\documentclass{IEEEcsmag}
\usepackage[numbers]{natbib}

\usepackage[colorlinks,urlcolor=blue,linkcolor=blue,citecolor=blue]{hyperref}
\expandafter\def\expandafter\UrlBreaks\expandafter{\UrlBreaks\do\/\do\*\do\-\do\~\do\'\do\"\do\-}
\usepackage{upmath}
\let\affil\relax
\usepackage{authblk}
\usepackage{listings}
\definecolor{IEEEblue}{RGB}{0, 112, 192}
\usepackage{booktabs}

\usepackage{algorithm}
\usepackage{algorithmicx}
\usepackage{algpseudocode}

\usepackage{amsmath}
\usepackage{amssymb}

\usepackage{microtype}

\usepackage{float}
\floatname{algorithm}{Prompt Strategy}

\DeclareRobustCommand*{\IEEEauthorrefmark}[1]{%
    \raisebox{0pt}[0pt][0pt]{\textsuperscript{\footnotesize\ensuremath{#1}}}}
    
\begin{document}

\sptitle{Theme Article: Inclusive Data Experiences}

\title{An Age-based Study into Interactive Narrative Visualization Engagement}

\author{
    \IEEEauthorblockN{
        Nina Errey\IEEEauthorrefmark{1}\textsuperscript{*},
        Yi Chen\IEEEauthorrefmark{2},
        Yu Dong\IEEEauthorrefmark{3}, 
        Quang Vinh Nguyen\IEEEauthorrefmark{4},
        Xiaoru Yuan\IEEEauthorrefmark{5}, 
        Tuck Wah Leong\IEEEauthorrefmark{6},
        and Christy Jie Liang\IEEEauthorrefmark{1}
    }
    \IEEEauthorblockA{\IEEEauthorrefmark{1}School of Computer Science, University of Technology Sydney, Sydney, New South Wales, Australia}
    \IEEEauthorblockA{\IEEEauthorrefmark{2}Beijing Key Laboratory of Big Data Technology for Food Safety, Beijing Technology and Business University, Beijing, China}
    \IEEEauthorblockA{\IEEEauthorrefmark{3}Advanced Interactive Technology and Application Laboratory, Computer Network Information Center, Chinese Academy of Sciences, Beijing, China}
    \IEEEauthorblockA{\IEEEauthorrefmark{4}Western Sydney University, Sydney, New South Wales, Australia}
    \IEEEauthorblockA{\IEEEauthorrefmark{5}Peking University, Beijing, China}
    
    \IEEEauthorblockA{\IEEEauthorrefmark{6}    School of Computing Technologies, RMIT University, Melbourne, Australia}

    \footnotesize{\textsuperscript{*}Nina Errey is the corresponding author.}
}

\markboth{THEME ARTICLE}{THEME ARTICLE}

\begin{abstract}Research has shown that an audiences’ age impacts their engagement in digital media. Interactive narrative visualization is an increasingly popular form of digital media that combines data visualization and storytelling to convey important information. However, audience age is often overlooked by interactive narrative visualization authors. Using an established visualization engagement questionnaire, we ran an empirical experiment where we compared end-user engagement to audience age. We found a small difference in engagement scores where older age cohorts were less engaged than the youngest age cohort. Our qualitative analysis revealed that the terminology and overall understanding of interactive narrative patterns integrated into narrative visualization was more apparent in the feedback from younger age cohorts relative to the older age cohorts. We conclude this paper with a series of recommendations for authors of interactive narrative visualization on how to design inclusively for audiences according to their age.

\end{abstract}

\maketitle

\chapteri{D}emographic characteristics such as an audiences’ age can impact how they interact and engage with digital media \cite{johnson2017designing, while2024gerontovis}. Interactive narrative visualization, or data storytelling is a form of digital media that has been shown to compel and explain information through an engaging end-user experience \cite{rogha2024impact, shi2022breaking, nowak2018micro}. It has thus been used to communicate critical topics such as the effects of climate change or election outcomes. Despite conveying important information, interactive narrative visualization is often designed without audience age in mind. One potential reason is that creating interactive narrative visualization is a labor-intensive process that requires expertise and creativity \cite{segel2010narrative}. Researchers have therefore concentrated on aiding authors to alleviate the challenges involved in interactive narrative visualization creation. Consequently, researchers have yet to fully explore the requirements of different audiences of interactive narrative visualization.

In this study we attempt to address a fundamental knowledge gap on the question of how audience age impacts engagement in interactive narrative visualization. We investigate if age significantly impacts engagement in interactive narrative visualization. Moreover, we qualitatively analyze why differences exist. We aim to give a concrete, evidence-based answer to whether authors of narrative visualization should prioritize audience age when designing interactive narrative visualization and how this should be done. 

To achieve our aim, we performed an empirical experiment. We developed three narrative visualization examples that employed different interactive narrative patterns tailored specifically for engagement \cite{bach2018narrative}. We randomly assigned 2400 participants to one of the three narrative visualization examples and measured their engagement using VisEngage \cite{hung2017assessing}. VisEngage is a self-reporting questionnaire specifically developed to measure engagement in visualization. The age groups were split into generations. A younger audience with an age of 18-27, a middle-younger audience between the ages of 28-43, an older audience of 44-59, and finally, the oldest age cohort consisting of over 60. These age groups were determined by generation boundaries. 

The results of our study revealed that there is a small but significant difference between older audiences and younger audiences’ engagement in narrative visualization. From our qualitative analysis, it was found that younger audiences were more observant of the interactive techniques employed that encouraged engagement. The terminology used by younger audiences was distinctly different from their older counterparts, where they described the cognitive processes involved in their interactive engagement. Older audiences were not so discerning and reported their negative engagement was partially due to the interactive device causing confusion or distraction. To our knowledge, this is the first study that investigates age groups and interactive narrative visualization engagement. This study contributes foundational research on interactive narrative visualization audiences. We conclude the work by presenting a series of recommendations for designing interactive narrative visualization inclusively based on our findings.

\section{Related Work}
\label{sec:related_work}

\subsection{Narrative Visualization}
\label{subsec:narrative_visualization}

Narrative visualization is a popular subject in the visualization community. Initial work on the topic was pioneered by Segel and Heer, who coined the term ‘narrative visualization’\cite{segel2010narrative}. They proposed a design space outlining genres and structures \cite{segel2010narrative}. Further foundational research on narrative visualization established an analytic framework by defining rhetorical techniques and possible transitions for story-sequencing \cite{hullman2013deeper,hullman2011visualization}. Recent advances in narrative visualization authoring processes include examples such as generative AI co-creation systems, visualization generation using natural language queries, and machine-guided workflows. 

While there is much research on the authoring process of narrative visualization, its impact on audiences is relatively under-explored. When viewing visualization it has been found that an audience’s personal beliefs impact their viewing experience \cite{peck2019data}. Furthermore, the story structure can influence end-user engagement in a narrative visualization \cite{liem2020structure}. More recently, a series of ‘narrative patterns’ were observed in narrative visualization, where the author's intent was correlated with a narrative device integrated into a data-driven story \cite{bach2018narrative}. Most have been studied and shown, with varying degrees of success, to encourage audience engagement \cite{rogha2024impact,shi2022breaking}.

No singular definition of interactive narrative visualization has been established. In this study, we use a broad, inclusive definition for interaction in visualization, as described by Dimara and Perin \cite{dimara2019interaction}. Their definition is thus, “Interaction for visualization is the interplay between a person and a data interface involving a data-related intent” \cite{dimara2019interaction}. To avoid being too vague, we refer to interactive narrative visualization, where an interactive modality such as scrolling, clicking, or inputting end-user-generated data is a key element of the narrative visualization. This differs from interactive narrative visualization from narrative visualization, such as data videos or data comics, where an interactive modality is not required. In the next section, we investigate engagement and how it is measured for visualization. 

\subsection{Engagement}
\label{subsec:engagement}

The human-computer interaction (HCI) community has long considered engagement a fundamental concept in user-centered design. Visualization research, in comparison, has relatively recently begun to seriously regard engagement \cite{hung2017assessing,mahyar2015towards}. The definition of engagement is often ambiguous and dependent on discipline. We adopt an HCI definition of engagement, which centers on the quality of the user experience and on the positive aspects of the interaction, particularly the phenomena associated with being captivated by technology. O’Brien et al. listed dimensions of engagement as including aesthetic appeal, novelty, perceived challenge, feedback, motivation, and affect \cite{o2008developing}. Alternatively, engagement can be viewed as a continuum from low to high \cite{mahyar2015towards}. Furthermore, engagement in narrative visualization has been viewed in the context of flow and fluid interaction \cite{mckenna2017visual}.

Numerous methods have been proposed to measure engagement in the field of visualization. Boy et al. evaluated engagement by analyzing time spent on interaction and user input \cite{boy2015storytelling}. Nowak et al. used elicitation interviews to examine factors including emotional affect and engagement in narrative visualization \cite{nowak2018micro}. A purpose-built method for measuring engagement in visualization was proposed by Hung and Parsons named VisEngage \cite{hung2017assessing}. VisEngage is a self-reporting questionnaire based on the user-engagement scale adapted for visualization \cite{hung2017assessing}. Similar adaptations were successful in other domains, such as social networking applications and games. A questionnaire comprising 22 questions, VisEngage addresses 11 engagement characteristics, where each characteristic corresponds to two questions.  VisEngage is a relatively robust method to measure end-user engagement in visualization \cite{hung2017assessing}.

\subsection{Age-based Research in Visualization}
\label{subsec:age-based_research}
Previous work on different age cohorts in visualization usually focused on either the very old or the very young. For example, visualization research with children has investigated pedagogical approaches for visual literacy and visualization design. Visualization research into elderly audiences has examined aspects of accessibility, comprehension and perception. The lack of understanding of the needs of different age cohorts is a known and cited dilemma in visualization research \cite{while2024gerontovis}.

In the HCI field, age-based research has found significant differences between age groups. For example, the time taken to perform input modalities of end-users compared to their age group found that older adults were significantly different from their younger counterparts. Although the results were inconclusive, strong evidence pointed to differences between older and younger age cohorts when using devices such as smart watches \cite{while2024dark}. 

In a report by the Interactive Advertising Bureau in the United Kingdom (UK) it was found that younger age cohorts find interactive advertisements more appealing than their older counterparts \cite{iab2024interactive}. However, in a report by the NN Group, unnecessary interactivity and flashy graphics are found to be ‘annoying’ by young adults \cite{nng2017youngadults}. The NN Group report explained that young adults are digital natives, who are “people raised in a digital, media-saturated world” and distinctly different to their older counterparts. Young audiences deemed ‘digital natives’ were more confident and less patient with user interfaces, according to the NN Group Report \cite{nng2017youngadults}.

Older adults are described as wary of technology. For example, older adults are supposedly less likely to prefer gamified user experiences and prefer text-based content. Moreover, older adults are reported to be less confident with user interfaces and are “hesitant to explore” \cite{nng2017youngadults}. Research into usability for older audiences is still lacking, where their preferences and behaviors are not adequately considered \cite{while2024gerontovis,nng2017youngadults}. While et al. introduced the term GerontoVis, which encapsulates data visualization design that primarily focuses on older adults, which they describe as a largely overlooked area of visualization research \cite{while2024gerontovis}. The contribution of this work is to provide evidence-based guidelines for inclusively designing interactive narrative visualization targeted toward an age group which will ultimately result in more effective interactive narrative visualization.

\section{Research Method}
\label{sec:research_method}

We conducted a crowd-sourced study using three narrative visualization examples as a stimulus to achieve our research aim of investigating if and how, end-user age impacts engagement in interactive narrative visualization. To see the interactive narrative visualization example code and raw data, please see the supplementary material \href{https://osf.io/6a5rp/?view_only=fc3f33f0407f4e19a14e8be6ca2e1526}{here}.

\begin{figure*}
\centerline{\includegraphics[width=36pc]{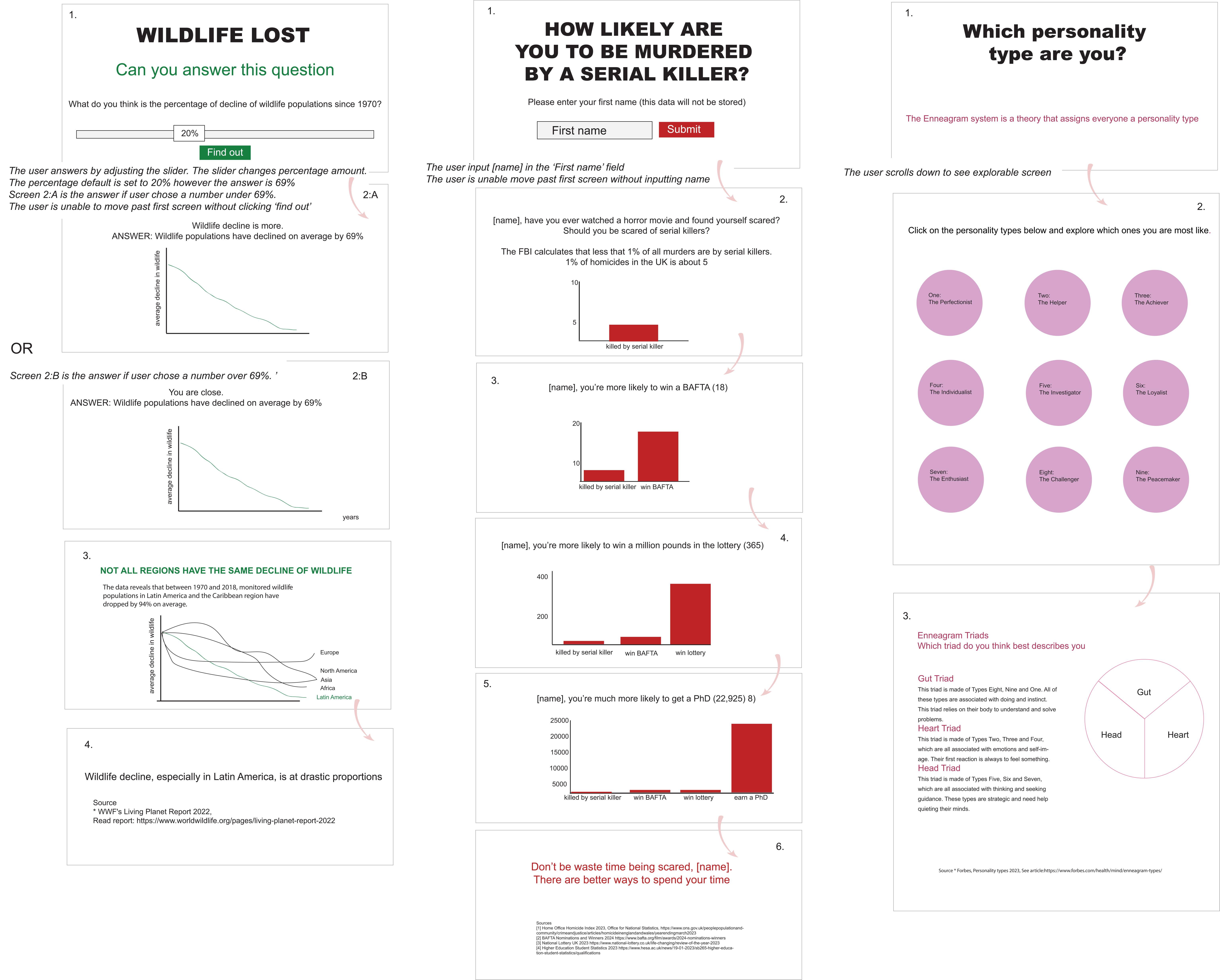}}
\caption{A diagram of interactive narrative visualization design examples A) Design One: ‘Make a Guess’ B) Design Two: ‘Breaking the Fourth Wall’ and C) Design Three: ‘Exploration’}\vspace*{-5pt}
\end{figure*}

\subsection{Experiment Design}
One of the primary challenges when designing an evaluation experiment is the dichotomy of localization and globalization. It is important to have results that can be globalized and, therefore, universally applicable. Conversely, it is necessary to have strict experiment parameters to report concrete results. To address this challenge, we developed three different narrative visualization examples that were similar in length but varied in topic. Each example is different in its data, messaging, and interactive narrative pattern. The intent of the integrated interactive narrative patterns was, however, similar – engagement \cite{hullman2013deeper}.

The research team iteratively developed three narrative visualization examples. Each example was inspired by publicly accessible interactive narrative visualizations from reputable publishers. Publishers that influenced our designs include The New York Times, ABC Australia, and The Pudding. By no means the only publishers of interactive narrative visualization, each of the aforementioned publishers is commended in online journalism awards for their interactive narrative visualization.

\subsubsection{\textbf{Design One: 'Make a Guess'}}

The first narrative visualization example, Design One, used a ‘Make a Guess’ interactive narrative pattern \cite{hullman2013deeper}. This pattern encourages engagement by stimulating the curiosity of an audience \cite{rogha2024impact}.

The audience is asked to guess an answer to a question, and the answer to the question is then revealed, affirming or disaffirming the accuracy of their answer. The objective of the ‘Make a Guess’ narrative pattern is that the audience questions their perception of reality by revealing a mismatch between perception and the actual data. An example of the ‘Make a Guess’ narrative pattern is a New York Times story on education titled ‘You Draw It: How Family Income Predicts Children’s College Chances.’  

In our study, the narrative visualization design example, which we refer to as Design One, ‘Make a Guess’, was based upon a dataset from the WWF's Living Planet Report 2022. It opened by asking the participant if they could answer this question; ‘What do you think is the percentage of decline of wildlife populations since 1970?’ Underneath the question was a sliding bar set by default to 20\% and a button stating, ‘find out.’ Once the participant clicked on the button, if the sliding bar was set to any number under 69\%, the participant would receive the same message - wildlife decline was more than their estimate. If they estimated above 69\%, they were answered with a ‘you are close.’ The default sliding bar amount, set at 20\%, was a deliberate ploy for the user to estimate a lower value, and thus be surprised by the correct answer. See Figure 1:A for a diagram of Design One.

\subsubsection{\textbf{Design Two: 'Breaking the Fourth Wall'}}

The second narrative visualization example, Design Two, used the ‘Breaking the Fourth Wall’ interactive narrative pattern. ‘Breaking the Fourth Wall’ is a term often cited in cinema and literature disciplines. In interactive narrative visualization, a direct question is asked of the audience, normally to input personal data. This creates a ‘self-story connection,’ which has been found to encourage engagement, as it includes the user within the story \cite{shi2022breaking}. The narrative visualization design example that inspired this study was a finalist in the online journalism awards. Published by the ABC Australia Story Lab, the narrative visualization is titled ‘See how global warming has changed the world since your childhood.’ In our study, Design Two broke the fourth wall by asking the user to input their name. Specifically, the user was asked to ‘please enter your first name (this data will not be stored)’ so that privacy concerns were availed with the assurance that data relating to the user’s name would not be stored. Design Two opened by stating, ‘How likely are you to be murdered by a serial killer?’ The user was asked to enter their first name click submit. Once submitted, a screen appeared with ‘[name], have you ever watched a horror movie and found yourself scared? Should you be scared of serial killers?’ This is followed by a basic bar chart explaining that '1\% of homicides in the UK are about 5.’ The narrative sequentially revealed itself as the user scrolled down the screen.’ All data sources are referenced at the end of the narrative visualization. See Figure 1:B for a diagram of Design Two.

\subsubsection{\textbf{Design Three: 'Exploration'}}

We refer to the third interactive narrative visualization design example as ‘Design Three.’ Differentiating from the previous two examples, Design Three integrated a narrative pattern that encouraged data exploration. The audience is asked to freely explore data so that they can create their narrative. Such an experience is described as a ‘reader-driven’ narrative visualization. Design Three was inspired by a narrative visualization that appeared in a digital publication called ‘The Pudding.’ The specific interactive narrative visualization was titled ‘A Visual Guide to the Aztec Pantheon’, which explained Aztec iconography. Similar in interface design to the Pudding example, Design Three encouraged users to click on the interface to explore information. The example asked users, ‘Which personality type are you? The Enneagram system is a theory that assigns everyone a personality type.’  The user was then asked to click on the personality types to find out which one they most like. See Figure 1:C for a diagram of Design Three.

\begin{figure*}
\centerline{\includegraphics[width=34pc]{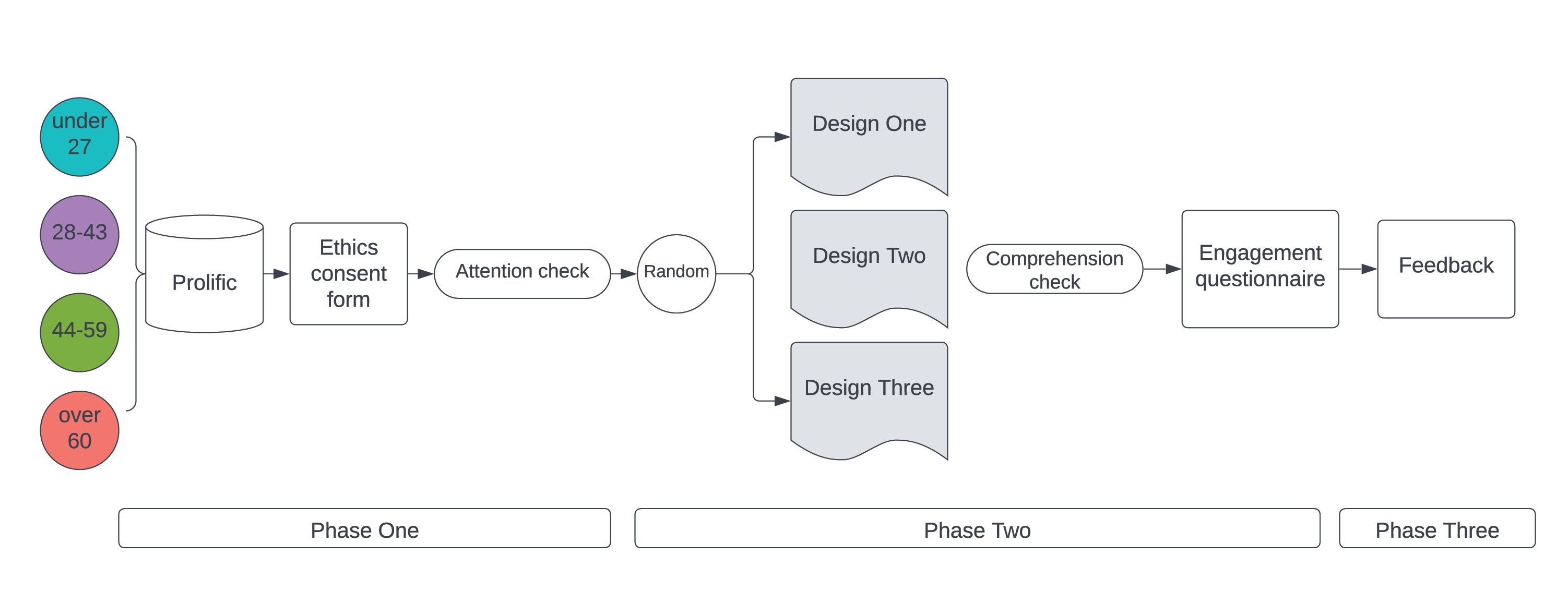}}
\caption{A flow diagram of experiment procedure. Experiment procedure: Prolific platform; ethics consent form; attention check, interactive narrative visualization designs; comprehension check; engagement questionnaire and feedback question.)}\vspace*{-5pt}
\end{figure*}

\subsection{Survey Design}

The survey instrument was adapted from the VisEngage engagement questionnaire \cite{hung2017assessing}. The survey instrument contained 22 questions, where the 11 engagement characteristics were allocated two questions each. For clarity, the wording of each question mentions a ‘data story’ rather than a narrative visualization. The participant could answer on a 7-point Likert scale from strongly disagree to strongly agree. Examples of the questions are as follows; ‘While using this data story, I found its look and feel to be pleasing’ or ‘The content or message of this data story was interesting to me.’ As described by Hung and Parsons, an overall engagement score can be achieved by adding together the results of each question \cite{hung2017assessing}. Strongly disagree is allocated a one and strongly agree is allocated a 7. Therefore, the maximum engagement score is 154, corresponding to high engagement, and the minimum is 22, corresponding to low engagement. 

\subsection{Experiment Procedure}
We conducted the experiment on the Prolific crowdsourcing platform. The experiment was in three phases. The first phase was where the participant exited Prolific and moved to the Qualtrics survey platform. They were asked to read and consent to the consent form, where ethics details were attached. In the next step, participants were asked for their Prolific ID, which was automatically inserted, and an attention check question. If the participant failed to consent, add their Prolific ID, or failed the attention check, their token was revoked, and they were returned to Prolific.

The second phase was where the participant was asked to ‘please interact with the data story and then answer the questions below.’ One of the three randomly allocated interactive narrative visualization designs was presented using an iFrame, where the interactive narrative visualization was hosted on an external server. An iFrame is an HTML element that allows you to embed another HTML document. After the iFrame, we posed a comprehension question to ensure participants had interacted with the narrative visualization. After the comprehension check question, the participant answered the 22 engagement questions. All questions were on a Likert scale, and all were mandatory.

The final phase of the experiment was a qualitative feedback question that asked, ‘Did you feel you were engaged in the data story? Why or why not?’ This question was not mandatory. The participant could then either submit a response or move to the next step, which returned them to the Prolific platform. See Figure 2 for flow chart of the experiment procedure.

\subsection{Participants}

We split each age cohort according to what are often described as ‘generations.’ Generational research is a foundational topic in social sciences; however, we would highlight that the labels used to describe generations can be loaded with stereotypical connotations. The objective of this research is not to perpetuate stereotypes associated with generational labels, and therefore we are not using the commonly used labels. We will refer to each age cohort by their age and refrain from using labels to diminish stereotypical connotations.
The youngest cohort consisted of ages ranging between 18-27. This age bracket saw participants born on or after 1997. Due to limitations with the crowd-sourcing platform, the youngest participant allowable age was 18. The second cohort had ages ranging between 28-43, where their birth year was on or between 1981 – 1996. This age cohort came to adulthood during the first years of the new millennium.  Born between 1965- 1980, this age cohort is between 44-59 years old. Finally, the oldest cohort consisted of participants with ages 60+ with a birth year on or after 1965. 

Our participant sample size was a result of a power calculation with a goal of a 95\% confidence level and a 4\% margin of error. The population was calculated based on the adult population size in the UK in 2022. Our ideal sample size was calculated at approximately 601 participants per age cohort, therefore, with three design examples with equally distributed participants with 200 in each group, our total ideal sample size was approximately 2400 participants. 

\section{Results}
\label{results}

\subsection{Hypothesis}

We expected to observe differences between the four age cohorts while factoring in the effect of the narrative visualization design examples. We firstly affirmed if a significant difference exists, specifically, our alternate hypothesis was as follows:

\textbf{H1:} There is a significant difference in engagement score and age cohort

\begin{table}
\caption{Combined engagement scores for all examples: mean and standard deviation (SD) per age cohort.}
\label{tab1}
\tablefont
\begin{tabular*}{17.5pc}{@{\extracolsep{\fill}}p{65pt}p{40pt}<{\centering}p{40pt}<{\centering}@{}}
\toprule
\textbf{Age Cohort} & \textbf{Mean} & \textbf{SD} \\
\colrule
18--27 & 112 & 15.8 \\
28--43 & 110 & 15.5 \\
44--59 & 109 & 17.1 \\
60+    & 109 & 16.8 \\
\botrule
\end{tabular*}
\end{table}

\begin{figure*}
\centerline{\includegraphics[width=32pc]{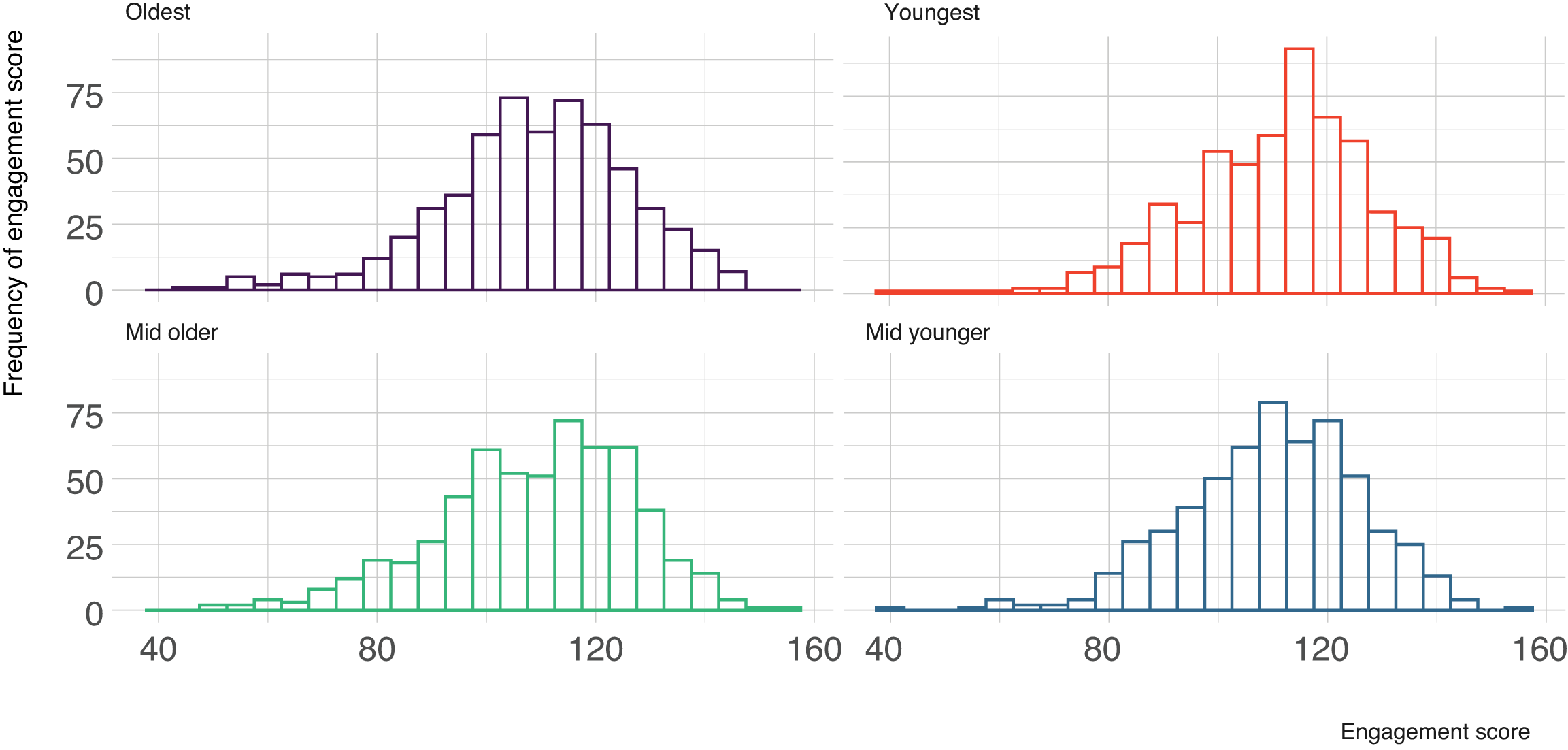}}
\caption{Series of histograms presenting each age cohort and frequency of engagement score.}\vspace*{-5pt}
\end{figure*}
\begin{figure*}
\centerline{\includegraphics[width=32pc]{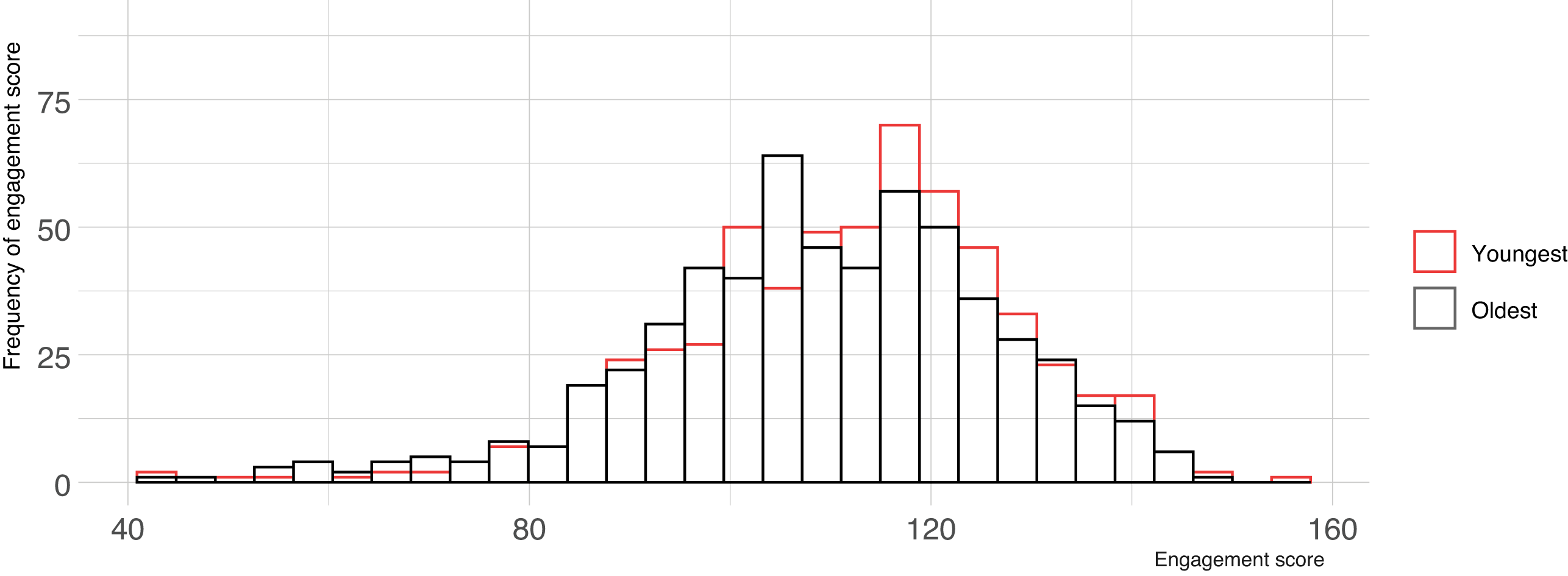}}
\caption{Histogram comparing oldest and youngest age cohort and frequency of engagement score on the same axis.}\vspace*{-5pt}
\end{figure*}

\subsection{Quantitative Analysis}
Out of the 2400 participants, 77 failed the comprehension check. To investigate which age cohort had the greatest engagement, our dependent variable needed to be the overall engagement score. This was calculated by adding together all question responses in the VisEngage questionnaire as recommended by Hung and Parsons \cite{hullman2011visualization}.

We normalized the engagement score data removing extreme outliers. Outliers were identified with engagement scores below 50, where we judged that their extreme response patterns indicated likely response bias. The number of outliers amounted to 43 participants, less than 2\% of participants. A Shapiro-Wilk test was performed, and the result was not significant. Approximately equal variances were tested using the engagement score as the dependent variable in Levene’s test, the result of which was not significant. The final number of participants totaled 2280 participants.  
 
We ran a one-way ANOVA to compare the effect of age and engagement score. The one-way ANOVA revealed that there was a statistically significant difference in engagement between at least two groups (F(3, 2426) = [2.81], p = [0.03]). Tukey’s HSD test for multiple comparisons found that the mean value of the engagement score was significantly different between the 60+ age cohort and 18-24 age cohort with a confidence interval of 95\% (p = 0.05, 95\% C.I. = [-5.1, 0.02]). All other age cohorts between groups means comparisons showed no significant differences. Figure 3 shows a series of histograms of engagement scores per age cohort for all combined design examples. Figure 3 illustrates that the 18-24 age cohort has a higher frequency of higher engagement scores relative to the 60+ age cohort. Figure 4 compares engagement scores of the 60+ age cohort and 18-24 age cohort, by overlaying them on the same axis and highlighting the difference. Table 1 shows the mean and standard deviation per age cohort. It further illustrates the small difference between the mean of each age cohort.

According to our test results we could accept our research hypothesis (H1). We then measured effect size of age on engagement score. The effect size of age cohort on engagement score, as measured by Cohen’s f, was 0.06, indicating a small effect size (95\% C.I. = [0.01, 0]). We analyzed the interaction effect of age cohort and narrative visualization design example on participant engagement scores. The two-way ANOVA revealed that there was no statistically significant interaction between the effects of age cohort and narrative visualization design example (F(6, 2355) = [1.52], p = [0.17). Therefore, when factoring in the narrative visualization design examples, they did not interact with participant engagement score. 

Finally, overall engagement in the interactive narrative visualization was positive for all age cohorts. The mean engagement score was 110 for all age cohorts combined. Furthermore, the median engagement score was 112. This data reveals that the majority of participants were positively engaged in interactive narrative visualization.

\subsection{Qualitative Analysis}
We investigated the thought processes of participants by analyzing their responses to a qualitative feedback question. The survey instrument asked, ‘Did you feel you were engaged in the data story? Why or why not?’ The aim was to shed light on the cognitive reasoning and reflections of participant on their engagement with the interactive narrative visualization. This question was not mandatory, and we received 2278 responses. To qualitatively analyze the data, we adopted an inductive theming approach using latent theming. We inductively coded lower-level themes determined by upper-level themes. Initially we determined upper-level themes by word frequency matched to potential engagement related issues. Examples such as ‘design’ and ‘interactive’ were deemed as upper-level themes according to their relative high frequency in the qualitative data. Two coders independently coded responses, and any inconsistencies were discussed.

\subsubsection{\textbf{Interactivity}}
‘Interactive’ was explicitly mentioned 7 times by the 60+ cohort. This was in contrast to the 18-27 age cohort, who explicitly mentioned ‘interactive’ or ‘interactivity’ 49 times. We analyzed the exact phrases that participants used in the 18-27 cohort relating to interactivity. Our analysis revealed that the younger audience attributed their engagement to interactivity, for example, “yes because it was interactive” or I was engaged as it was an interactive task” (both comments from 18-27 age cohort, Design One).
The interactive device in Design Two, aimed to include the audience in the story and thus encourage engagement. We coded 22 instances where the 18-24 age cohort recognized that this was the aim of the interactive device in Design 2. For example, “Yes as it was interactive and by using my name felt personal” (18-27 age cohort, Design Two). In contrast, we coded 3 instances in the 60+ age cohort specifically mentioning the interactive device in Design Two. 
We found further evidence of perceptiveness in Design Three from the 18-27 age cohort, where we coded 18 responses that noticed the interactive device, which in this case required clicking and exploring the interface. For example, “I did feel I was engaged in the data story as I had to click to find the information as well as scroll for more information” (18-27 age cohort, Design Three). The goal of the interactive device in Design One, ‘Make a Guess’ was to illustrate a mismatch between audience expectation and reality. We coded 17 responses from the 18-27 age cohort that referenced the interactive device. For example, ‘The decline is higher’ which made me feel engaged” (18-27 age cohort, Design One). In contrast the 60+ age cohort mentioned the interactive device in Design One 6 times. While we observed that 18-24 age cohort was relatively more aware of the interactive devices this does not mean other age groups were oblivious, only their perceptive feedback was less frequent. For example, “I felt engaged as there was an interactive question where I could enter what I thought to be the answer. This made the impact of learning the true answer heavier as I was engaging with the story” (44-59 age cohort, Design One) or “I found the personality types interesting flip over to read and to associate the descriptions with the images you gave for the personality” (60+ age cohort, Design Three).  
When recognized by the older age cohorts, there were 17 instances where the interactive device had the opposite effect of encouraging engagement and reported as a distraction or confusion,  “I was distracted by, initially, not realizing I had to scroll down the box to gain more information” (60+ age cohort, Design Two) or “I found the chart near the end distracting as it seemed confusing.” (44-59 age cohort, Design Three). In the 18-27 age cohort, there were 2 observable reports of the interactive devices causing confusion and none causing distraction. The two reports were varied on the reasoning for why the interactive narrative visualization was deemed confusing. 
The younger age cohorts did report that they preferred an easy-to-use interface, where the interactive device did not detract from the user experience. We coded 57 responses from the younger age cohorts that mentioned interface functionality, in both positive and negative light. As described here, “it would have been better if you did not have to scroll down” (28-43 age cohort, Design One), or it “felt like a bit of a gimmick, why not just have the information under the pictures instead of needing to click nine times” (18-27, Design Three).  The data suggests that regardless of the audiences’ age, usability influences engagement. For the older age cohorts, however, the effects of poor usability cause greater effect than a minor aggravation, but feelings of confusion and distraction.

\subsubsection{\textbf{Cognition}}
In older audiences, it has been shown that complex visualizations can be cognitively demanding, requiring users to remember and interpret multiple information pieces simultaneously. We found evidence that older audiences expressed a preference for less complex visualization, for example, “I don't find it particularly easy to interpret graphs or charts and I loathe Venn diagrams, so, I possibly had to concentrate more than other participants in order to ensure that I was interpreting the information correctly” (60+ age cohort, Design Two) or “I found the charts a little confusing to begin with - possibly my age!” (44-59 age cohort, Design One). 
Furthermore, older audiences mentioned that they were required to revisit the narrative visualization to fully comprehend it. We coded 9 instances where participants from the older age cohorts mentioned they missed data in the interactive visualization, this was compared one instance that appeared in the 28-43 age cohort and none were observable in the 18-27 age cohort. “My biggest problem with it was the need to scroll down. At first I didn't realize there was more data below” (60+ age cohort, Design One). This reveals older audiences might miss crucial information if the representation is too complex or the interactive device is not clearly marked.  

\subsubsection{\textbf{Aesthetic Appeal}}

We examined the responses of the participants who reported they were not engaged. The primary reason, reported by the 18-27 age cohort, was criticism of the aesthetic appeal of the narrative visualizations. For example, Design One, had a black background, which was described as ‘dated’, where for example it was stated, “It felt quite outdated especially with the colors” (18-27, Design One).  Design Three, was described as ‘cluttered’, for example, “No as the text for the personality types 1-9 was cluttered” (18-27, Design Three). We coded 12 instances in the 18-27 age cohort commenting on the color palette of Design One. Comparably, the 60+ age cohort mentioned the color palette of Design One 3 times. These comments reveal the importance of sound aesthetic design, particularly for an interactive narrative visualization aimed at a younger audience. 

One difference observed between the older and younger age cohorts was their preference toward text integrated into the narrative visualization. Younger audiences preferred less text that was divided into smaller sections which they could control, for example “I felt that the gradual reveal of information meant that it was easier to compartmentalize statistics and different pieces of information rather than looking at a solid block of text, it felt more intuitive” (18-27, Design One). We found 14 instances where the 18-27 age cohort mentioned that they preferred text was gradually revealed where the end-user could control the pace. 
Older audiences asked for more information that provided context to the interactive narrative visualization. For example, “It was thought provoking and I felt it needed more pages to explain what has been lost and why” (60+ age cohort, Design One). 6 instances were coded of the 60+ age cohort asking for more information. The 18-27 age cohort had one observable instances where they asked for more information. 

Finally, we coded 1437 strongly positive responses. This is in line with our quantitative data analysis, where overall positive engagement was reported for all age cohorts, however, in line with our quantitative analysis, slightly less for the older age cohorts. The 60+ age cohort reported 326 instances of positive engagement, 338 by the 44-59 age cohort, 382 by the 28-43 age cohort, and 391 times by the 18-27 age cohort. This reflected a generally positive opinion of engagement in the narrative visualization examples across all age cohorts.

\section{Discussion and Future Work}
\label{subsec:discussion}
The primary aim of this study was to find out if and how audience age impacts engagement in interactive narrative visualization. This research is a fundamental step toward giving greater credence to the needs of the audience when designing interactive narrative visualization. We established that audience age did have a small impact on end-user engagement in interactive narrative visualization. The data revealed the greatest difference was between the 60+ age cohort and the youngest 18-27 age cohort. 

In this section we present a series of recommendations for designing interactive narrative visualization for specific age cohorts. 

\subsection{Designing Narrative Visualization for Older Audiences}

\subsubsection{Older audiences are more attuned to usability difficulties.}

We found that the 60+ age cohort reported a slightly lower engagement score compared to the 18-27 age cohort. We investigated the qualitative data to find out why their engagement was lower. We found 17 instances where the 60+ age cohorts reported feeling distracted and confused by the interactive narrative patterns integrated into the narrative visualization. These negative reactions could explain their lower engagement scores of the older age cohort compared to the youngest age cohort. 
It is important to note that all age cohorts desired ease of use. The primary difference between age cohorts was the extent of the negative reaction to usability-related concerns. For example, older age cohorts reported not recognizing functionality in the interactive narrative visualization, such as scrolling. We found 9 instances where it was stated that data was missed due to usability difficulties in the 60+ age cohort. Missing crucial data contained in an interactive narrative visualization might result in the central messaging being misconstrued.
Younger age cohorts recognized that they were required to scroll but preferred that it was not required. The findings of this study highlight the importance for narrative visualization authors to prioritize usability for all audiences, however, especially if the visualization is aimed at an older aged audience. Moreover, for older age cohorts, important information contained in an interactive narrative visualization might be missed if the interactive device or data representation is too complex. 

\subsubsection{Older audiences desire the option for more information.}
It was found that older audiences asked for more information and context to the narrative presented. We suggest that authors of narrative visualization provide the option for the end-user to access further information about the presented topic. While it might not be necessary for primary messaging contained within the narrative visualization, more text provides context to older audiences, where assumed knowledge might not be present. 

\subsection{Designing Narrative Visualization for Younger Audiences}
\subsubsection{Younger audiences understand interactive narrative patterns.}
Qualitative data analysis indicated a stark difference in the terminology used by the youngest age cohort relative to the older age cohorts. Interactivity was mentioned 49 times by the youngest age cohort, and usually in a positive light. Furthermore, the youngest age cohort seemed to have a perceptive understanding of how the interactive narrative pattern achieved its intent of encouraging engagement. For example, the younger age cohort’s responses to Design Three, ‘Exploration’ specifically outlined how exploring the data through clicking was more engaging than simply reading it. Our study sheds light on the depth of understanding of interactive devices held by the youngest age cohort, illustrated by their descriptive feedback. 
The results of this study do not disprove the NN Group report, which stated it was a myth that young adults “crave multimedia and innovative design.”\cite{nng2017youngadults} Rather, the results of this study give credence to the fact that young adults are accustomed and therefore more understanding of interactive devices. Increased engagement is, therefore largely due to the young adults’ ability to perceive the intent of the author. Whereby recognizing that as an audience, younger age cohorts are expected to be engaged, therefore they are engaged. For future authors of interactive narrative visualization targeted towards younger age cohorts, it is recommended to use interactive narrative patterns with explicit intent. Duplicitous or superfluous use of interactive narrative patterns would likely be recognized and thus could result in lowered engagement. 

\subsubsection{Younger audiences appreciate aesthetics.}
Positive aesthetic appreciation is a known contributor to engagement \cite{o2008developing}. Aesthetics were reported to directly impact engagement for the younger age cohorts and was attributed as their primary reason for not engaging in the narrative visualization. Criticism regarding design was overwhelming prevalent in this age cohort. These criticisms included negative feedback on color, imagery and layout. We observed that younger audiences preferred an interface that was less cluttered and thus easy to digest. It is recommended that authors of narrative visualization segment their information thoughtfully. Furthermore, while it is helpful to use automated tools for narrative visualization generation, the role of the narrative visualization author continues to be of importance to aesthetically evaluate the overall design and flow of the narrative visualization.

\subsection{Design Recommendations for All Audiences}
\subsubsection{Interactive narrative visualization has a broad appeal.}
One positive outcome of our study is the apparent appeal of interactive narrative visualizations. The mean engagement score for all age groups combined was 110.  The positive mean average indicates overall positive engagement in interactive narrative visualization. In addition, the qualitative analysis evidenced a largely positive reaction, where 63\% of feedback responses reported a strongly positive engagement. This finding evidences that as a communication medium, interactive narrative visualization can engage a broad audience. Authors of interactive narrative visualization should not shy away from designing narrative visualization for older audiences. The study presented here shows that, when designed inclusively, interactive narrative visualization is an engaging medium for all age demographics. 

\subsection{Future Research Opportunities}
The sheer volume of data that was generated by the large participant base in this study requires greater inspection. Nuanced differences in the data were not adequately considered as they were beyond the scope of this work. For example, we have simply added the engagement 22 questions, where each of the 11 engagement characteristics received two corresponding questions, in the VisEngage questionnaire to achieve a final engagement score. We added them together as it was recommended by Hung and Parsons \cite{hung2017assessing}. However, it could be interesting to investigate if individual characteristics appeal to age groups differently. Furthermore, it stands to reason individual differences in design examples impact the engagement of audiences. We have provided the raw data from the experiment in the supplementary material, where we encourage future researchers to inspect and analyze the data in greater detail. 
As the study of narrative visualization audiences is a relatively emerging area of research, we suggest other demographics that could be worthy as a focus of investigation. Other demographics divided by education, visual literacy, technological literacy or location could prove to be interesting avenues of investigation. The empirical evidence reported in this study was written with the goal of making conclusions over a broad audience divided merely by age cohort, however it is clear age is just one of many audience characteristics that can potentially impact engagement. 

Ultimately this work results in more effective interactive narrative visualization as it can better inform future interactive narrative visualization design and research. Empirical visualization research can overlook the age of their participant base, this study shows that age can impact experiment results and should be reported \cite{while2024gerontovis}. We hope the findings of this study encourages authors and researchers to seriously consider their audience when designing or researching interactive narrative visualization in the future, where age is but one of many audience characteristics that should be considered. 

\section{Limitations}
There are several limitations to this study. The foremost limitation is that there are but three narrative visualization design examples. Optimally, we would have used a multitude of examples. However, the scope of this study dictated a limit of three. It should be noted, however, that each design example uses one of the three interactive narrative patterns that are described for engagement \cite{hullman2013deeper}. The objective of this study was a broad approach, where we used differing topics, narrative patterns, and designs. It is unfeasible to study all possible combinations of topics, narrative patterns, and designs. We believe that the three examples we developed were adequate to achieve our study’s aims. 

Another notable limitation is that this study is only representative of an audience based in the UK. The availability of the oldest age cohort from countries outside the UK and the US was specifically challenging and disappointing to the international research team. The uneven distribution of older participant country locations resulted in a decision to focus the experiment on participants from the UK. Rather than a skewed result, we prefer our results to concretely representing the behaviors of peoples from one locale. Furthermore, the premise of our study is to question whether different demographics engage differently with narrative visualization; therefore, it stands to reason that the locale of participants might impact study results.  Comparing audience engagement across multiple countries is outside the scope of this work.

Our study’s participant pool was recruited from an online crowd-sourcing platform. Recruiting participants from an online platform necessitated a level of technical proficiency from participants. Furthermore, the study required participants undertake the experiment on a desktop computer. These factors resulted in our study’s participant pool being skewed toward more technically proficient participants. Future work could consider recruiting participants from offline sources where a lack of technical proficiency might impact the study outcomes.

For future researchers, we have provided our designs and code from the interactive narrative visualization examples in the supplementary material. We encourage researchers to replicate this study using varied design examples, alternative languages or using alternative recruitment strategies.

\section{Conclusion}
\label{sec:conclusion}
To communicate effectively, content authors are required to recognize the needs, preferences, and behaviors of their intended audience. The outcomes of this study suggest that audience age impacts their engagement in interactive narrative visualization. Older audiences that are in the 60+ age cohort find that interactive narrative patterns integrated into narrative visualization cause usability difficulties. Younger age cohorts do not experience the same response when presented with interactive narrative patterns. Younger age cohorts recognize and appreciate interactive narrative patterns and are thus more engaged than their older counterparts. Our results lead to valuable implications for designing future interactive narrative visualization, where we encourage authors to give greater consideration to their audience when designing interactive narrative visualization.

% \section{ACKNOWLEDGMENTS}

\def\refname{REFERENCES}

% \bibliographystyle{IEEEtran}
% \bibliography{main}

% \begin{IEEEbiography}{{\textcolor{IEEEblue}{Nina Errey}}} is [] at []. Contact her at [].
% \end{IEEEbiography}

% \begin{IEEEbiography}{{\textcolor{IEEEblue}{Yu Dong}}} is assistant professor at Computer Network Information Center, Chinese Academy of Sciences at Beijing, China. Contact him at dongyu@cnic.cn.
% \end{IEEEbiography}

\end{document}